\documentclass[onecolumn]{elsart3p}

\usepackage{graphics}
\usepackage{graphicx}
\usepackage{dcolumn}
\usepackage{bm,amsmath,amssymb,amsfonts}
\usepackage{psfrag}

\newcommand{\be}{\begin{equation}}
\newcommand{\ee}{\end{equation}}
\newcommand{\ba}{\begin{eqnarray}}
\newcommand{\ea}{\end{eqnarray}}

\begin{document}

\begin{frontmatter}



\title{\vspace{-1.2cm}
\vspace{5mm}%
A new Parameterization for the Pion Vector Form Factor}


\author{C.~Hanhart}
\address{ Institut f\"ur Kernphysik, Institute for Advanced Simulation 
and J\"ulich Center for Hadron Physics,\\
Forschungszentrum J\"ulich, D--52425 J\"{u}lich, Germany}

\begin{abstract}
A new approach to the parameterization of pion form factors is presented and
for illustration applied to the pion vector form factor. It has
the correct analytic structure, is at low energies consistent with recent high
accuracy analyses of $\pi\pi$ scattering phase shifts and, at high energies,
maps smoothly onto the well--known, successful isobar model.
\end{abstract}

\begin{keyword}
Pion form factor \sep Omn\`es representation
\sep 11.55.Bq  
\sep 13.40.Gp  
\end{keyword}
\end{frontmatter}

\section{Introduction}

In recent years the knowledge about the low energy two--pion system has
improved significantly, both experimentally as well as theoretically:
for the low partial waves phase shift parameterizations of high accuracy
exist from different dispersive analyses, either involving data
only~\cite{madrid_new}, or involving both data as well as constraints from
chiral symmetry~\cite{bern}. The analyses are based on Roy or Roy-type 
equations that respect analyticity as well as crossing symmetry. Especially,
left--hand cuts are included without approximation.

In contrast to this, pion form factors or production reactions are often
modeled either by sums of Breit-Wigners or improved versions
thereof~\cite{GS,HL,CGK} or by the K--matrix formalism.
In case of overlapping resonances unitarity gets violated by the former 
ansatz.
The K--matrix provides a clear improvement compared to the Breit-Wigner
parameterization, since two--body unitarity is built in. However, in general
analyticity is violated. On the one hand, in the standard treatment not the
full dispersive corrections are considered (in the expressions for the self
energies only the imaginary parts and their analytic continuation are being
kept and not the full expressions --- c.f. Eq.~(\ref{Ldef}) below), although
some works include them (see, e.g., Ref.~\cite{anisovich2011}), on the other
hand, the left hand cuts are not treated properly ---  if they are included
at all, in
order to fit the scattering amplitudes, they are often in the same way
included in the production amplitude although there the left hand cuts are
different or, as in case of form factors, even absent. 

To be specific, in this work we focus on form factors and scattering with the
goal to present simple formulas that allow for a data analysis that is
consistent with analyticity and unitarity, however, without the necessity to
solve complicated integral equations. In addition, we present formulas that, by
construction, in the low energy regime map smoothly and consistently onto what
can be derived from the high accuracy analyses mentioned above and thus for
the scattering even include the proper left hand cuts.

As an example and for demonstration we apply the formalism in this paper to
the pion vector form factor, related to $\pi\pi$ scattering in the
$p$--wave. The experimental situation for the $\rho$ resonances beyond the
$\rho(770)$ is at present not very clear: different experiments find
indications for different resonances. While two resonances ($\rho(1450)$ and
$\rho(1700)$) appear to be well established, 3 additional resonance candidates
can be found in the literature
 --- for a summary of the current
situation see 'note on the $\rho(1450)$ and the $\rho(1700)$' in the Review of
Particle Physics~\cite{PDG}. The
formalism presented here could be an important step forward to clarify the
situation for it allows for a simultaneous, consistent analysis of various
channels/observables.

As an example and for demonstration we apply the formalism in this paper to
the pion vector form factor, related to $\pi\pi$ scattering in the
$p$--wave. The experimental situation for the $\rho$ resonances beyond the
$\rho(770)$ is at present not very clear: different experiments find
indications for different resonances --- for a summary of the current
situation see 'note on the $\rho(1450)$ and the $\rho(1700)$' in the Review of
Particle Physics~\cite{PDG} --- to parameterize the vector form factor beyond
$s=1$ GeV$^2$ all studies agree on the need to include at least two resonances
in addition to the $\rho(770)$, which is elastic.
 The formalism presented here could be
an important step forward to clarify the situation for it allows for a
simultaneous, consistent analysis of various channels/observables.
  The number of parameters needed for each resonance agrees to
standard parameterizations. The advantage of the parameterization presented
here is that we do not need to approximate the left hand cuts, the consistency
with low energy phase shift is ensured by construction and the connection
between scattering and form factors is properly implemented.

The paper is structured as follows: the most important formulas are motivated
and presented in Sec.~\ref{sec:sum}. Their more detailed derivation is given
in Sec.~\ref{sec:derivation}.  Results are presented in Sec.~\ref{sec:results}
and the paper closes with a short summary in Sec.~\ref{sec:summary}. The
two-potential formalism which forms the basis for the derivation is introduced
in the Appendix.

\section{Summary of most important results}
\label{sec:sum}

\begin{figure}[t]
\begin{center}
\psfrag{T}{$\tilde T$}
\psfrag{TR}{$t_R$}
\psfrag{VR}{$V_R$}
\psfrag{V}{$\tilde V$}
\psfrag{G}{$\Gamma$}
\psfrag{g}{$G$}
\psfrag{L}{$\Sigma$}
\includegraphics*[width=0.7\linewidth]{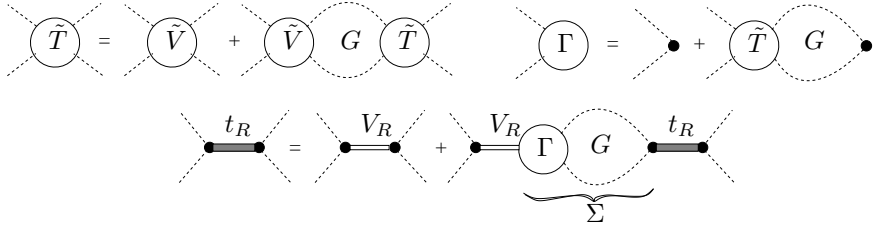}
\caption{Diagrammatic representation for the various ingredients of the
formalism. 
 \label{fig:diags}}
\end{center}
\end{figure}

The fundamental quantity in the current analysis is the pion vector form
factor $F_V(s)$ defined via
\begin{equation}
\langle \pi^+(q_1)\pi^-(q_2)|J^\mu|0\rangle = (q_1-q_2)^\mu F_V(s) \ ,
\label{fdef}
\end{equation}
where $s=(q_1+q_2)^2$. For clearity in this paper we will only discuss
 $F_V(s)$, although the formalism is more general.

Below the first inelastic threshold only the two--pion channel needs to be
considered.
Then  it is possible to give a closed form expression for the form factor $F_V$
solely in terms of the elastic scattering phase shift --- the so--called Omn\`es
solution~\cite{Omnes:1958hv}.  It is derived from a dispersion relation using the fact that
\begin{equation}
\mbox{disc}(F(s))=2i\sigma T(s)^*F(s) \ ,
\label{ima}
\end{equation}
where $T$ denotes the
on--shell elastic scattering amplitude, and 
$\sigma=\sqrt{1-4m_\pi^2/s}/(48\pi)$ the two--body phase space.
Here 'disc' denotes the discontinuity of the form factor
defined via
$$
\mbox{disc}(F_V(s))=F_V(s+i\epsilon)-F_V(s-i\epsilon)=2i{\rm Im}(F) \ .
$$
If the elastic phase shift is $\delta(s)$, with
\begin{equation}
T(s)=\frac1{\sigma} \sin (\delta(s))e^{i\delta(s)} \ ,
\label{telast}
\end{equation}
 then
the form factor reads in the absence of bound states
\begin{equation}
\label{elastFdef}
F(s)=\Omega[\delta](s)P_A(s)
\end{equation}
with the Omn\`es function
\begin{equation}
\Omega[\delta](s)=
\exp\left\{ \frac{s}{\pi}\int_{4m^2}^\infty \frac{ds'}{s'}
\frac{\delta(s')}{s'-s}\right\} \ .
\label{omnes}
\end{equation}
where
the function $P_A(s)$ is a polynomial. Its degree may be fixed by the large $s$ behavior
of the form factors. For values of $s<1$ GeV$^2$ this methodology was used by
various authors for the pion vector form factor, see, e.g. Refs.~\cite{HL,gassermeissner,guerrero,pich,yndurain,oller,ours}.

In this letter a formalism is derived that 
\begin{itemize}
\item[$\bullet$] smoothly maps onto Eqs.~(\ref{telast}) and (\ref{elastFdef}) for the
  elastic $T$ matrix and the form factor respectively at low energies;
\item[$\bullet$] is an analytically improved version of the well known
 $K$--matrix approach at higher energies;
\item[$\bullet$] can be presented in closed form.
\end{itemize}
Especially Eqs.~(\ref{telast}) and (\ref{elastFdef}) 
will be replaced by 
\begin{equation}
T(s)_{ij}=  \delta_{ij}\delta_{1i}\tilde T(s) + T_R(s)_{ij} = \delta_{ij}\delta_{1i}\tilde T(s) + \xi_i\Gamma_{\rm
  out}(s)_i t_R(s)_{ij}\Gamma_{\rm in}(s)^\dagger_j\xi_j \ ,
\label{Tsplit}
\end{equation}
and
\begin{equation}
F(s)_i = \Gamma_{\rm out}(s)_{i}\left[ 1 - V_R(s)\Sigma(s)\right]^{-1}_{ik}M_k \ ,
\label{FFstructure}
\end{equation}
with 
$$
\Gamma_{\rm out}(s)_{i}=\Gamma_{\rm in}(s)^\dagger_i = \left\lbrace
{{\Omega[\tilde\delta](s) \mbox{ for } i=1} \atop {\ \ \ 1   \qquad \mbox{ otherwise}}}\right. 
$$
and
\begin{equation}
\tilde T(s)=\frac1{\sigma_1} \sin (\tilde \delta(s))e^{i\tilde \delta(s)} \ ,
\label{ttildeelast}
\end{equation}
\begin{equation}
t_R(s)_{ij} = \left[ 1 - V_R(s)\Sigma(s)\right]^{-1}_{ik}V_R(s)_{kj}
\label{tRdef}
\end{equation}
and
\begin{equation}
\Sigma_i(s) = \frac{s}{\pi}\int_{s_{\rm thr }{} i}^\infty \frac{ds'}{s'}
\frac{\sigma_i(s')\xi_i^2\left|\Gamma_i(s')\right|^2}{s'-s-i\epsilon} \ .
\label{Ldef}
\end{equation}

The resonance potential $V_R$ as well as the production vertex $M$ will be
discussed in detail in the next section.
The functions $\xi_i$ ($\sigma_i$)
 parametrize the centrifugal barrier (phase space) operative in channel $i$ --- the concrete
 parametrization used will be given in the next section. 
In Fig.~\ref{fig:diags} a graphic representation of the various quantities 
is given.

Eq.~(\ref{FFstructure}) is the central result of our paper. If all vertex
functions were chosen to be constant, it would reduce to the famous
$P$--vector formalism~\cite{ian}. However, since in our case $\Gamma_1$, which
enters explicitly in the expression for $F_1$ as well as through $\Sigma_1$, is
non--trivial, Eq.~(\ref{FFstructure}) provides a generalization to the
conventional treatment.

The compared to Eq.~(\ref{telast}) additional contribution to the $T$ matrix, $T_R$, will be constructed
such that it gets negligible at low energies, such that
$$
T(s)_{ij} \simeq  \delta_{ij}\delta_{1i}\tilde T(s)  \mbox{ for } s<1 \  \mbox{GeV}^2 \ ,
$$
and thereofore we may identify $\tilde \delta(s)$ with the high accuracy phase shifts
derived from, e.g., Roy equations mentioned in the introduction --- see also 
Sec.~\ref{sec:results}. At the same
time the non-analytic pieces embodied in $\left[ 1 - V_R(s)\Sigma(s)\right]^{-1}_{ik}M_k$
get heavily suppressed with the result that Eq.~(\ref{FFstructure}) maps onto
Eq.~(\ref{elastFdef}) --- the net effect of this term below $s=1$ GeV$^2$ is
a 10\% increase in the pion radius compared to what comes from
Eq.~(\ref{omnes}), as demanded by the data.
In the next section the equations are  derived. Results are
be presented in Sec.~\ref{sec:results}.

\section{Derivation of the Formalism}
\label{sec:derivation}

Eqs.~(\ref{ima}) and (\ref{omnes}) apply only, if the interactions are purely
elastic --- for the latter it is even necessary that they are elastic up to
infinite energies. Clearly this is not realistic. However, experimental data
show that at higher energies inelasticities are typically accompanied by
resonances. We therefore split the full, partial wave projected interaction potential $V$ into
two pieces
\begin{equation}
V(s)_{ij} = \tilde V(s)_{ij} + V_R(s)_{ij} \ , 
\label{vdef}
\end{equation}
where $i$ and $j$ denote the channels. The crucial feature for this approach
is that the potential $\tilde V$ needs to be specified at no point. All what
is needed are the corresponding phase shifts $\tilde \delta$.
We now $postulate$ the following properties:
\begin{itemize}
\item[$\bullet$] the potential $\tilde V$ is purely elastic, such that 
$\tilde V$ is non--vanishing only for $i=j=1$; 
\item[$\bullet$] deviations in the $\pi\pi$ phase--shifts from $\tilde \delta$
 come solely from
$s$--channel resonances;
\item[$\bullet$] all long ranged forces (driven by left--hand cuts) are in the elastic $\pi\pi$ interactions
  $\tilde V$; all interactions in other channels are regarded as short ranged. 
\end{itemize}
These are the model
dependent assumptions of this approach (note, in case of the traditional isobar model one needs to assume
that $all$ interactions are mediated by $s$--channel resonances). Based on these Eqs.~(\ref{ima}) and (\ref{omnes})
can be easily generalized to multiple channels providing a convenient
parameterization for both scattering as well as production amplitudes.

\begin{figure}[t]
\begin{center}
\includegraphics*[width=0.8\linewidth]{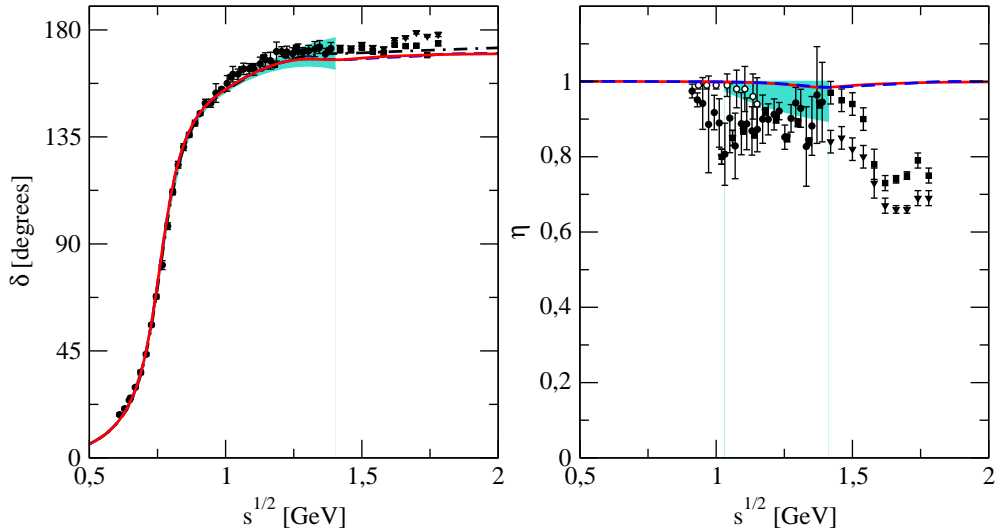}
\caption{Fits result for the pion $p$--wave phase shift (left panel) and
 inelasticity (right panel). The red solid band (blue dashed lines)
  denotes the result of the fit \#2 (\#1).
The dot--dashed line in the left panel refers to the input phase $\tilde \delta$. 
 Data are from Ref.~\cite{hyams} (solid
  dots --- only data below 1.4 GeV are shown~\cite{WO}), Ref.~\cite{hyams1} (solid squares for solution $(---)$; solid triangles for solution $(-+-)$), 
and Ref.~\cite{propo} (open dots) and Ref.~\cite{madrid_new} (turquoise band).
 \label{fig:phases}}
\end{center}
\end{figure}

The full scattering $T$--matrix appears as the solution of a Bethe--Salpeter
equation
with input potential $V$ defined in Eq.~(\ref{vdef}).
Using the two potential formalism (see Ref.~\cite{twopotform} and Appendix) it is straightforward to
derive the
decomposition given in Eq.~(\ref{Tsplit}).

The resonance potential may be written as (note: not all resonances are in $V_R$; elastic
resonances may be included in $\tilde T$ --- as the $\rho(770)$ in the example
below)
\begin{equation}
\bar V_R(s)_{ij} = -\sum_{l=1}^n g_i^{(l)}G_{(l,l')}g_j^{(l')} \ ;
  \ \ G_{(l,l')}=\frac{1}{s-m_{(l)}^2}\delta_{(l,l')} \ ;
\ \ V_R(s) = \bar V_R(s)- \bar V_R(0) \ . 
\label{VRdef}
\end{equation}
In Fig.~\ref{fig:diags} a graphic representation of the various quantities 
is given.
The potential is subtracted once at $s=0$ to ensure that the phase of the full $T$
matrix at low energies agrees to the input phase $\tilde \delta$.
Clearly, the procedure does not guarantee a priori that the phase of the full
$T$--matrix is close to that of $\tilde T$ in the whole range where $\tilde T$
is well determined, however, in practice this is indeed the case: 
 as can be seen from the left panel of 
Fig.~\ref{fig:phases}, at energies below 1 GeV all curves shown, including that
for the input phase shift, are indistinguishable.
In this sense a single subtraction is sufficient for the  consistency of
the approach and therefore higher subtractions are not necessary. 

The $M_i$ denote point like
source terms for the production of particles into channel $i$. It may be
written
as
\begin{equation}
M_k = c_k - \sum_{l=1}^n {g_i^{(l)}G_{(l,l')}\alpha^{(l')}s} \  .
\label{Mdef}
\end{equation}
Note, as a consequence of gauge invariance we use a photon--resonance
coupling linear in $s$\footnote{On the Lagrangian level this means a coupling of
the resonance to the photon field via $F^{\mu\nu}\partial_\mu V_\nu$ with
$F^{\mu\nu}$ for the electro--magnetic field strength tensor and $V_\nu$ 
for the resonance field.}. This coupling at the same time suppresses the
influence of heavier resonances on the low $s$ region.
The parameters $c_k$ allow for a direct transition of the photon to the 
different continuum channels. In case of the pion
vector form factor, discussed in detail below, charge conservation demands
$c_1=1$,
the other $c_k$ are free parameters of the fit. This is in contrast to
the isobar model, where commonly only resonant interactions are allowed,
however, these kinds of couplings appear naturally in more microscopic
approaches~\cite{maurice,freds,leupold,mauricenew}.


The centrifugal barrier factor in the two--pion channel is given by
\begin{equation}
\xi_1=\sqrt{s-4m_\pi^2}^{L_1}\left(\frac{\lambda^2}{\lambda^2+s}\right)
\label{xidef}
\end{equation}
 with $L_1$ for the angular momentum of the pion pair.
The final factor on the right hand side is introduced to tame the growth of
 $\xi_1$ that comes from the centrifugal barrier factor. In this work
$\lambda$ is not treated as a parameter determined by the fit, but we will
vary its value within some ranges and use the variation in the fit results as
some estimate for systematic uncertainties. Note: all
channels discussed explicitly here have $L_i=1$, however, e.g. the $\rho f_0$
channel couples to the $\pi\pi$ vector channel in an even partial wave.

To parameterize the inelastic channels we use for $i=2$ a structureless 4$\pi$
channel via  the phase space factor $\sigma_2=\sqrt{1-16m_\pi^2/s}^7/(48\pi)$, which provides the proper
scaling of the four--body phase space near the threshold. For the barrier
factor we use $\xi_2= \sqrt{s-16m_\pi^2}[\lambda^2/(\lambda^2+s)]$.
Although the $\bar KK$ channel contributes significantly to isoscalar $\pi\pi$
interactions, it gives negligible contributions in the isovector
channel~\cite{Simon}.
Therefore  
 we include as additional inelastic channel ($i=3$) the $\pi\omega$ channel. For
a discussion on the possible role of the $\pi\omega$ channel on the
$\pi\pi$ inelasticity see Ref.~\cite{costa77,bastianneu}. 
We take $\sigma_3=\sqrt{(s-(m_\omega+m_\pi)^2)(s-(m_\omega-m_\pi)^2)}/s/(48\pi)$ and $\xi_3=
48\pi\sigma_3 s[\lambda^2/(\lambda^2+s)]$.

\begin{figure}[t]
\begin{center}
\includegraphics*[width=0.8\linewidth]{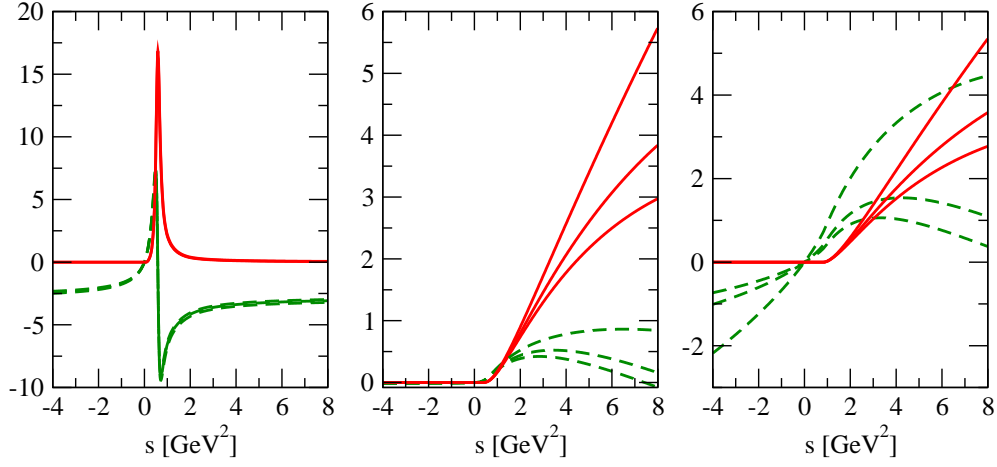}
\caption{Real and imaginary parts for the self energies $\Sigma_i(s)$
defined in Eq.~(\ref{Ldef}). The
solid (dashed) lines show the imaginary (real) part.
The left, middle and right panel refer to the self energies
from the 2$\pi$, 4$\pi$ and $\pi\omega$ channels. The different curves
refer to different values of the cut--parameter $\lambda$ in the vertex
functions
as described in the text.
 \label{fig:selfenergies}}
\end{center}
\end{figure}

 It is straightforward to show that the imaginary parts of
the self--energies read (see Appendix) 
\begin{equation}
\mbox{Im}(\Sigma_i(s))=\sigma_i \xi_i^2|\Gamma_i(s)|^2 \theta[s-s_{{\rm thr }{} i}] \ ,
\label{imL}
\end{equation}
where $\theta[...]$ denotes the step function equal to 1 (0) for positive
(negative) arguments.
Thus the self energies can be calculated from a properly subtracted dispersion integral.
Since $\Sigma(s_0)$ at some $s_0$ can be absorbed
into the resonance masses we
here use a once subtracted version --- c.f. Eq.~(\ref{Ldef}).  
 The resulting self energies $\Sigma_1(s)$, $\Sigma_2(s)$ , and $\Sigma_3(s)$
 are shown in left, middle and right panel of Fig.~\ref{fig:selfenergies},
 respectively.  In all panels the solid lines refer to the imaginary parts
 while the dashed lines refer to the real parts.  The rise of the imaginary part of both $\Sigma_2$ and
 $\Sigma_3$ comes from the centrifugal barrier terms $\xi_2$ and $\xi_3$,
 respectively, its curvature comes from the vertex function 
$\lambda^2/(s+\lambda^2)$.
 Each panel contains three
 curves of each kind: the most strongly curved ones come from $\lambda=4$ GeV,
 the least curved one from $\lambda=6$ GeV, while the middle one referes to
 $\lambda=5$ GeV.

To calculate the form factor we may write (using the symbolic notation of the Appendix)
\begin{equation}
\xi{ F} = \xi M + TG \xi M \ ,
\label{Fdefeq}
\end{equation}
 where $G_i$ is the operator representation for the integration over all
intermediate $n$--body states of channel $i$ and the production vertices $M_i$
were defined in Eq.~(\ref{Mdef}).
Note the appearance of the centrifugal barrier factors in Eq.~(\ref{Fdefeq}).
They appear, since the discontinuity equations are derived on the basis of the
full transition current that, in addition to the form factor, contains a
kinematic factor for pion pairs in a partial wave higher than $s$--wave ---
see Eq. (\ref{fdef}). 
Inserting Eq.~(\ref{Tsplit}) into 
Eq.~(\ref{Fdefeq}), we get, after deviding by $\xi_i$
$$
{ F_i} = M_i+\tilde T_{ij}G_j(\xi_j/\xi_i)M_J + (T_R)_{ij}G_j(\xi_j/\xi_i)M_j = \Gamma_{{\rm
  out } i}(\delta_{ij}+(t_{R})_{ij}\Gamma_{{\rm
  in } j}^\dagger G_j \xi_j^2 )M_j \ .
$$
To proceed we may use the definition of the self energy,
$\Sigma_i=\Gamma_{{\rm  in } i}^\dagger G_i\xi_i^2$, to write
$$
t_R\Sigma = \left[1-V_R\Sigma\right]^{-1} V_R\Sigma = -1+\left[1-V_R\Sigma\right]^{-1}  \ .
$$
Here we needed to assume that the range of interactions in the production
vertex and in the vertex functions of the resonances is similar in all
channels,
for only then the same loop integral $\Sigma_i$ can be used as self energy
contribution for the resonances as well as convolution integral  
of $M_i$ and the resonance potential.
From this  we get
$$
{ F} = \Gamma_{\rm out} \left[1-V_R\Sigma\right]^{-1}M \ ,
$$
which agrees to Eq.~(\ref{FFstructure}).

It is important to observe that the expression given in
Eq.~(\ref{FFstructure}) is consistent with the coupled channel version of 
the unitarity relation for form factors, Eq.~(\ref{ima}), since

\begin{eqnarray}
\nonumber
\mbox{disc}({ F_1}) &=& \mbox{disc}(\Gamma_{{\rm out } 1})\left[1-V_R\Sigma\right]^{-1}_{1j}M_j 
+\Gamma_{{\rm out} 1}^*\mbox{disc}(\left[1-V_R\Sigma\right]^{-1})_{1j}M_j \\ \nonumber
&=&2i\tilde T^* \sigma_1 \Gamma_{{\rm out } 1}\left[1-V_R\Sigma\right]^{-1}_{1j}M_j 
+\Gamma_{{\rm
  out } 1}^*(\left[1-V_R\Sigma^*\right]^{-1}V_R)_{1k}\mbox{disc}(\Sigma)_k\left[1-V_R\Sigma\right]^{-1}_{kj}M_j
\\ \nonumber
&=& 2i\left( \tilde T^*\delta_{1k}+\xi_1\Gamma_{\rm
  out}^*{}_1{(t_R)}{}_{ik}\Gamma_{\rm
  out}^*{}_k\xi_k\right)\sigma_k (\xi_k/\xi_1) {\Gamma_{{\rm
    out} k}\left[1-V_R\Sigma\right]^{-1}_{kj}M_j}  \\ 
&=& 2iT^*_{1k}\sigma_k (\xi_k/\xi_1)F_k
\end{eqnarray} 
where in the intermediate step the unitarity relation for the vertex function,
Eq.~(\ref{ima}), and the self energy, Eq.~(\ref{imL}), were used.

An interesting observable is the ratio $r$ of the total cross section for
$e^+e^-$
annihilation into hadronic states with $I=1$ other than $\pi^+\pi^-$ over
$\sigma_{e^+e^-\to \pi^+\pi^-}$ --- a
compilation of this quantity can be found in Ref.~\cite{Simon}.
In this ratio the unitarization effects in the
resonance $T$--matrix cancel largely. For example, in case of only one
inelastic channel we get
\begin{equation}
r = \left|\left(\frac{\xi_2\sigma_2\Gamma_{{\rm out}\ 2}}{\xi_1\sigma_1\Gamma_{{\rm
        out}\ 1}}\right)\frac{(1-V_{R \, 11}\Sigma_1)M_2+V_{R \, 12}\Sigma_1 M_1}
{(1-V_{R \, 22}\Sigma_2)M_1+V_{R \, 12}\Sigma_2M_2}\right|^2 \ ,
\end{equation}
clearly being very directly sensitive to the resonance parameters.

\begin{figure}[t]
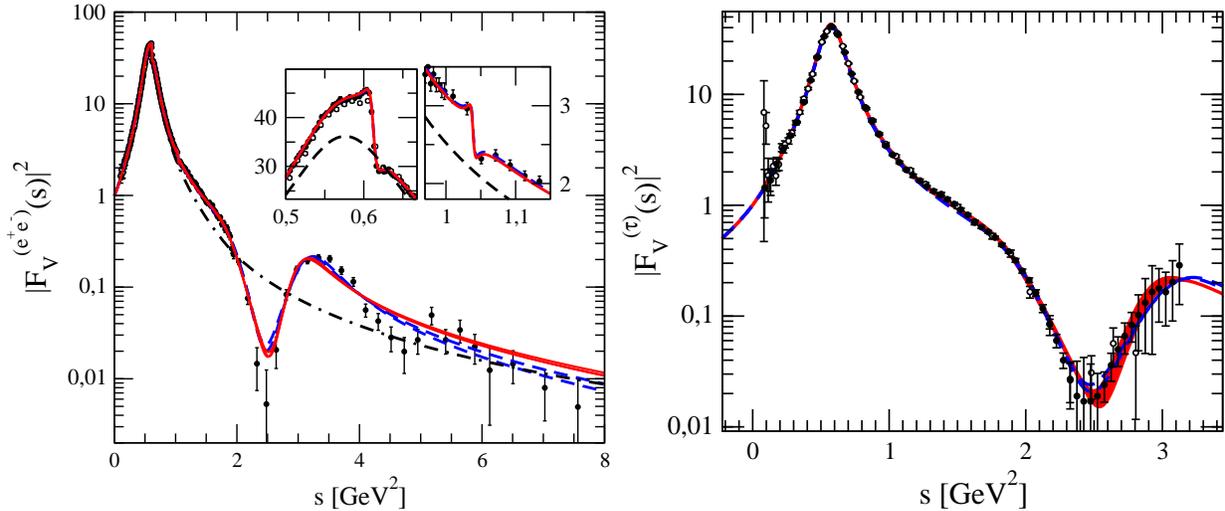

\begin{center}
\includegraphics*[width=0.49\linewidth]{pionFF_fitcomp.eps}
\includegraphics*[width=0.49\linewidth]{pionFF_fitcomp_taudata.eps}
\caption{Fit result for the pion vector form factor $F_V$. Left panel: For the
  neutral
channel --- the left (right) inlay shows a zoom to the region where
$\rho$-$\omega$ ($\rho$-$\phi$) mixing is visible.
 The red solid band (blue dashed lines)
  denotes the result of fit \#1 (\#2). The black
dot--dashed line shows the form factor derived from the Omn\`es function only.
 Data are from
the reaction $e^+e^-\to \pi^+\pi^-$ presented in Refs.~\cite{babarFF,KLOE}.
Right panel: For the charged channel. Data are from Belle~\cite{BelleFF} and CLEO~\cite{cleo}. 
 \label{fig:pionFF}}
\end{center}
\end{figure}

\section{Inclusion of isospin violation}

The vector form factor in the two pion channel is directly accessible from two
reactions, namely from $e^+e$--annihilation and from $\tau$ decays. In the
former
case, in addition to what was discussed so far,  isospin violating
mechanisms need to be included.

There are two well established effects --- $\rho-\omega$ and $\rho-\phi$
mixing. Both are visible  as   narrow structures in the pion vector form factor located at the masses
of the $\omega$ and $\phi$, respectively (c.f. inlays in Fig.~\ref{fig:pionFF}). The inclusion of these mixings in the present
formalism is straight forward --- we here use a slightly modified version
to what is used in Ref.~\cite{yndurain}, namely, for the neutral, $\pi^+\pi^-$, channel
\begin{equation}
c_1 \longrightarrow
c_1 \ \! \left(1+\kappa_1\frac{s}{s-m_\omega^2+im_\omega\Gamma_\omega}+\kappa_2\frac{s}{s-m_\phi^2+im_\phi\Gamma_\phi}\right) . 
\end{equation}
We here use $m_\omega=0.7826$ GeV, $\Gamma_\omega=0.0085$ GeV, $m_\phi =
1.0193$ GeV, and $\Gamma_\phi= 0.0043$ GeV.
 The strength
parameters $\kappa_i$ are part of the fit. 

In Refs.~\cite{freds} it was pointed out that in the form factor
for the neutral channel the mixing with the photon needs to be taken into account as well.
In Refs.~\cite{freds} only the mixing with the $\rho(770)$ was
considered. Here the photon can also mix with the higher resoncances.
The complete effect can be included via
\begin{equation}
\bar V_R^{EM}(s)_{ij} = -\sum_{l,l'}^n \hat g_i^{(l)}G_{(l,l')}^{EM}\hat g_j^{(l')} \ ;
\ \ V_R(s)_{ij} = \bar V_R(s)_{ij}- \bar V_R(0)_{ij}-\frac{e^2c_ic_j}{s} \ . 
\label{VRdefEM}
\end{equation}
and
\begin{equation}
M_k = c_k - \sum_{l=1}^n {\hat g_i^{(l)}}{G_{(l,l')}^{EM}\alpha^{(l')}s} \  ,
\label{MdefEM}
\end{equation}
with
\begin{equation}
\hat g_i^{(l)} = g_i^{(l)}-e^2 \alpha^{(l)} c_i  \ .
\label{EMcorrs}
\end{equation}
The resonance propagators $G^{EM}$ come from mixing of the hadronic resonance
basis with the photon and are defined via
\begin{equation}
G^{EM}=\left[1-G \ s\alpha \alpha^T\right]^{-1}G \ ,
\end{equation}
where the bare propagators $G$ were defined in Eq.~(\ref{VRdef}). 
Thus the mixing with the photon does not introduce any new parameter.
For simplicity we do not include the additional isospin violating
effects listed in Ref.~\cite{freds}.

\section{Results for the pion phases, inelasticities and form factors in the $p$-wave}
\label{sec:results}

All parameters introduced --- $\kappa_i$, $c_i$, $g_i^{(l)}$, $\alpha^{(l)}$, $m_{(l)}$
--- are real as long as all (relevant) channels are treated explicitly as a
consequence of time reversal invariance.  In case of $N$ channels, for
each explicit resonance we need to include $N+2$ real parameters. 
 This number
agrees exactly to what is needed, e.g., in the $K$ matrix formalism.
In addition there are the $N-1$ direct couplings of the photon to
the continuum channels, $c_i$ --- charge conservation fixes $c_1=1$.  For the
predominantly elastic resonance $\rho(770)$, which is included via $\tilde
\delta$, the only free parameters are the mixing paramters
$\kappa_1$ and $\kappa_2$. 
Denoting the number of resonances included explicitly by $n$ (c.f. Eq.~(\ref{VRdef})), we thus have
$[(n+1)(N+2)-1]$ parameters in total.
In this paper we discuss two classes of fits: one including 2 resonances and
2 channels, the other including 2 resonances and 3 channels. Thus,
in the former case we have 11 in the latter 14 parameters to be adjusted to
the data.

The resonance parameters will be determined by a fit to data on the time--like pion vector
form factor as well as inclusive data on inelastic channels.
All fits were performed using the MINUIT package of the CERN library.
As input we need the elastic phase shifts $\tilde \delta$, which
largely fix the properties of the $\rho(770)$.

For the input phase shifts $\tilde \delta$ we will use for energies below $s_{\rm cut}=1.4^2$ GeV$^2$ the
central
values for the phase shift
provided in Ref.~\cite{madrid_new} --- see Eqs. (A7) and (A8) therein. For energies above this value we smoothly
extrapolate the phase shift to a value of $\pi$ via
\begin{equation}
\tilde \delta (s)=\pi + (\tilde \delta(s_{\rm
  cut})-\pi)\left(\frac{\Lambda^2+s_{\rm cut}}{\Lambda^2+s}\right) \ .
\end{equation}
It turns out that for $\Lambda\geq 2$ GeV the results in the time like region are basically
insensitive to the actual value used~\footnote{This is correct up to an
  unphysical
pole located at $s=-\Lambda^2$, which, however, does not influence visibly the
amplitude for $s>-\Lambda^2$.}. We thus chose $\Lambda = 10$ GeV
in what follows.
The asymptotic value $\pi$ for the phase ensures that the vertex function
$\Gamma_1(s)$
decreases as $1/s$ as demanded for the vector form factor.
The resulting elastic $\pi\pi$ phase shifts  are
shown as the black dot--dashed line
in the left  panel of Fig.~\ref{fig:phases}. The form factor from the Omn\'es
function alone is shown as the black dot--dashed line 
 in Figs.~\ref{fig:pionFF}. It provides an acceptable description
of the data  up to 
$s=1$ GeV$^2$, although there are some deviations visible (see inlays in 
the left panel). At higher energies a significant deviation becomes visible.

\begin{table}
\begin{center}
\begin{tabular}{| c|| c| c| c| c|| c| c| c | c | c |  c | c | c| c | c |  }
\hline
fit & $\kappa_1 \times 10^3$ & $\kappa_2 \times 10^3$ & $c_2$ & $c_3$  & $m_{(1)}$ & $m_{(2)}$  & $g^{(1)}_1$  & $g^{(1)}_2$  & $g^{(1)}_3$  &
$g^{(2)}_1$  & $g^{(2)}_2$ & $g^{(2)}_3$ & $\alpha^{(1)}$  & $\alpha^{(2)}$   \\ 
\hline
\#1A & -2.2(1) & -0.6(1) & -4.9(2) & -- & 1.53(1) & 2.3(1) & -1.40(3) &
-38(1) & -- & 1.9(1) & 114(10) &  -- & -0.04(1) & 0.39(1) \\
\#1B & -2.2(1) & -0.5(1) & 4.5(1) & -- & 1.52(1) & 2.4(1) & 1.38(2) & 
-36(1) & -- & -1.8(1) & 120(1) & --   & 0.04(1) & -0.36(1) \\
\hline
\#2A & -2.2(1) & -0.6(1) & 3.8(1) & -6.2(4) & 1.57(1) & 2.0(1) & 1.6(1) & 
-50(3) & 2.8(7) & 1.1(1) & -43(7) & 14(2) & -0.01(20)   & 0.73(7) \\
\#2B & -2.2(1) & -0.6(1) & 3.6(1) & -5.8(4) & 1.58(1) & 2.0(1) & -1.6(1) & 
51(2) & -1.4(6) & 0.9(1) & -37(5) & 11(1) & 0.01(10)   & 0.77(7) \\
\#2C & -2.1(1) & -0.5(1) & 3.6(3) & -5.5(4) & 1.58(2) & 2.0(1) & -1.6(1) & 
50(3) & -0.8(6) & 0.8(1) & -37(5) & 11(1) & -0.01(10)   & 0.8(1) \\
\hline
\end{tabular}
\caption{Fit parameters for the various fits.
The uncertainties listed refer to the
  statistical uncertainty of the individual fits only.
Masses and $\alpha_i$ are given in GeV and GeV$^{-2}$, respectively; all other
couplings are dimensionless.
Fit \#1A and \#2A refer to $\lambda=4$ GeV, \#1B and \#2B to 5 GeV and \#2C to 6 GeV,
as described in the text.\label{tab:parameters}}
\end{center}
\end{table}

\begin{figure}[t]
\begin{center}
\includegraphics*[width=0.7\linewidth]{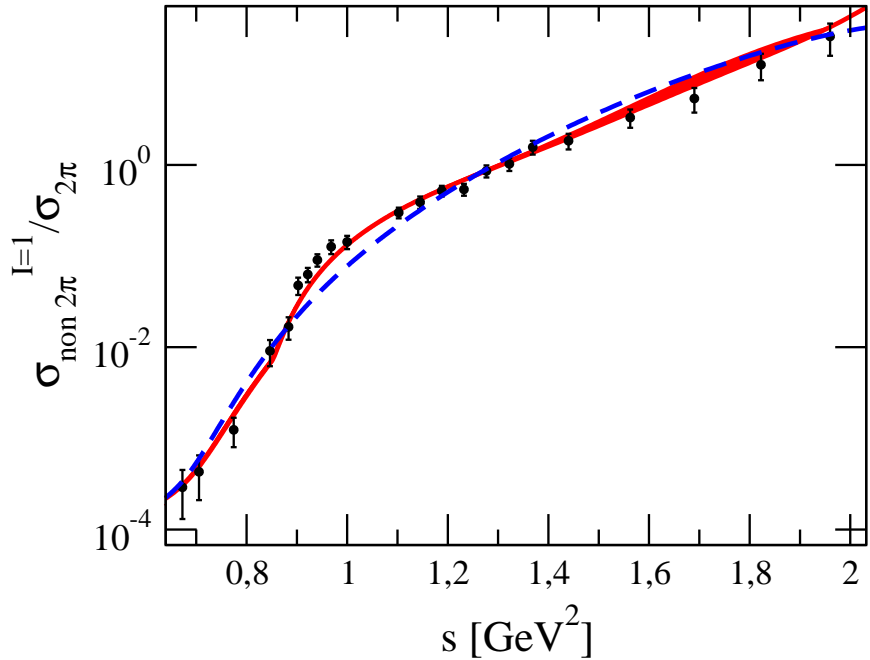}
\caption{Results for the ratio $r=\sigma_{e^+e^-\to ({\rm non 2 }
\pi)}^{\rm I=1}/\sigma_{e^+e^-\to \pi^+\pi^-}$. 
 The red solid band (blue dashed lines)
  denotes the result of the fits \#2, (\#1).
 Data are from the compilation of Ref.~\cite{Simon}. 
 \label{fig:FFnon2pi}}
\end{center}
\end{figure}

The pion vector form factor shows pronounced structures for $s>1.2$ GeV$^2$
(c.f.  Fig.~\ref{fig:pionFF}). These call for the inclusion of at least two
resonances.  Resonances come with pronounced phase motions that are, however,
not visible in the $\pi\pi$ phase shifts
(c.f. Fig.~\ref{fig:phases}). Therefore at least one inelastic channel needs
to be included, for then this phase motion can appear in the phase of the form
factor only; see discussion below. For this channel we choose a structure less
4$\pi$ channel --- the corresponding fits we call Fit \#1.  It turns out that
in order to get the overall scale of the cross section ratio $r$ right
a  non--vanishing value for the parameter $c_2$, defined in Eq. (\ref{Mdef}), is needed.  In order to
investigate the model dependence of the parametrization we performed fits with
different cut--off parameter $\lambda$ in the vertex functions
(c.f. Eq.~(\ref{xidef})).  If the dynamics considered explicitly in the model
is sufficient to describe the physics in the energy range studied, one would
expect a very mild dependence of the calculated observables on $\lambda$. This is indeed what is found when
$\lambda$ is varied between 4 and 5 GeV, however, for larger values of
$\lambda$ a simultaneous fit to $F_V$ and $r$ was not possible. The blue
dashed lines in Figs.~\ref{fig:phases}, \ref{fig:pionFF} and
\ref{fig:FFnon2pi} for phases and inelasticities, the pion vector form factor
from $e^+e^-$ annihilation and $\tau$ decays, and the ratio $r$, respectively,
show the fit results for $\lambda=4$ (fit \#1A) and 5 (fit \#1B). The
corresponding parameters are given in Table~\ref{tab:parameters}. 

Since the allowed range of variation of $\lambda$ was limited and since the
energy dependence of the ratio $r$ is not well described with only one
inelastic channel, we performed a second group of fits with the $\omega\pi$
channel included in addition. Now $\lambda$ can be varied over a larger range
--- the red band in Figs.~\ref{fig:phases}, \ref{fig:pionFF} and
 \ref{fig:FFnon2pi} reflects the variation of the fit results when $\lambda$
is changed from 4 GeV (fit \#2A) to 6 GeV (fit \#2C) --- even for $\lambda=10$ GeV a reasonable
fit can be found with slightly larger $\chi^2$. 
The parameters extracted from the fits are listed in Table~\ref{tab:parameters}. 

  In case of the pion vector form factor we fit to the
 BaBar data on $e^+e^-$ annihilation only, for it extends to higher
 energies~\footnote{In this exploratory study we regard this as appropriate
   since in this work we do not perform an uncertainty estimate of the
   parameters extracted.}.
 Especially, neither the space like form factor data nor the
 $\tau$ data were included in the fit --- the result comes out as a
 prediction.  The parameters determined in both fits are given in
 Tab.~\ref{tab:parameters}.  It is important to note that for the pion vector
 form factor alone both fits are of similar quality: for fit \#1 and \#2 we
 have $\chi^2/d.o.f$=1.2 and 1.4, respectively.  However, only fit \#2
 provides an acceptable description for $r$.  This result nicely illustrates
 that one should not analyze the pion vector form factor without looking at
 the non 2$\pi$ channels at the same time. Note that at $s=2$
 GeV$^2$ Ref.~\cite{Simon} reports a value of $r\sim 26$, which shows that
 already at this relatively low energy the $2\pi$ channel provides only a
 small fraction of the $e^+e^-$ annihilation rate in the $I=1$ channel.

The $s$ dependence of the non 2$\pi$ data shown in Fig.~\ref{fig:FFnon2pi}
 calls at least for two inelastic channels, since there is  
a change in slope visible at around $s\sim 0.9$. In fit \#2 this is accounted
for by the inclusion of the $\pi\omega$ channel.
However, even this three channel fit is still too simplified, for there should
be not only correlations amongst the 4 pions in channel 2 included, e.g.
from $a_1\pi$, $\rho\rho$, and $\rho \sigma$. In addition there are also
channels like $\eta\pi\pi$. The data included in the
current study does not allow one to disentangle these and therefore to improve
the description of $r$ further. What is necessary is an inclusion of the large number
of exclusive measurements available from $e^+e^-$ annihilation. We leave this to a future study.

In contradistinction to most approaches in this model direct couplings of the
photon to the continuum channels are included; all fits called for non--zero
values of these (c.f.  Tab.~\ref{tab:parameters}). However, the same effect
could also be achived by inclusion of a heavy ($m>4$ GeV) resonance. In order
to decide if the dynamics included here is sufficient a fit to more, exclusive
data is compulsory. It is important to stress that the mass parameters given
in the table are bare parameters that get renormalized by the self
energies. It is therefore possible that the lowest resonance poles come out
very close in both fits, for the unitarization effects are different.
However, we postpone the determination of pole positions, which requires the
evaluation of the elastic $T$--matrix $\tilde T$ in the complex plane, to a
later work.

A direct $\rho\pi\omega$ coupling was shown to be significant, e.g.,
in~\cite{GMSW}. The information how strong this transition is in the present
approach is contained in the values of the corresponding residue at the
$\rho(770)$ pole, however, to
get access to this also calls for an analytic continuation of the amplitudes
into the complex plane.

\begin{figure}[t]
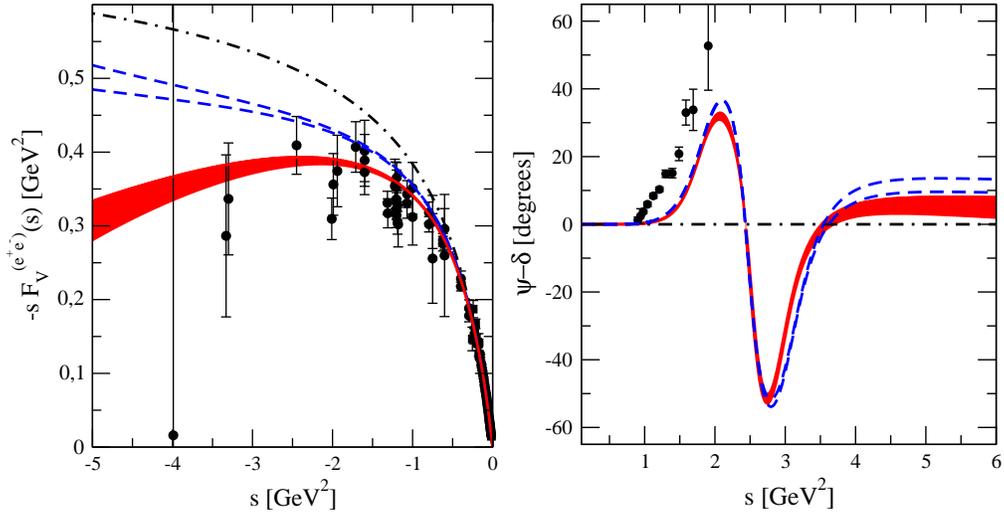

\begin{center}
\includegraphics*[width=0.4\linewidth]{piFF_spacelike.eps}
\includegraphics*[width=0.4\linewidth]{phasedifference.eps}
\caption{Left panel: Fit result for the pion vector form factor at space--like
energies. 
The lines are the same as in Fig.~\ref{fig:pionFF}. Data are from Ref.~\cite{Na7FF,spacelikecollection}.
Right panel: phase difference between $\psi$, the phase of the form factor,
and the scattering phase shift $\delta$. The data shown indicate the upper
bound of the phase shift difference presented in Ref.~\cite{Simon}.
 \label{fig:pionFFsmalls}}
\end{center}
\end{figure}

The results of the two fits for the vector form factor at space--like energies
are shown in the left panel of Fig.~\ref{fig:pionFFsmalls}. Both lead to a pion radius slightly
enhanced compared to what comes from the Omn\`es function itself, shown by the
dot--dashed line: through the inclusion of the high-lying resonances the 
 mean square charge radius of the pion increases by nearly 10\% from
0.40 fm$^2$ to 0.44 fm$^2$. The latter value is consistent
with the values extracted in Refs.~\cite{eta2pipigamma,gilberto}.
For the curvature we find $c_V=3.9$ GeV$^{-4}$, in line with Refs.~\cite{ours,ana}.
 The effect of the
higher resonances is quantitatively  in line with expectations from
dimensional analysis that predicts an effect on the mean square radius of order of the square 
of the inverse resonance mass $\sim 0.02$ fm$^2$.
At higher space like energies the results of both fits are consistent with the largely model independent 
bounds for the form factor derived in Ref.~\cite{ananew}, while
the form factor from the Omnes function is inconsistent.

The left panel of Fig.~\ref{fig:phases} nicely illustrates that within the
formalism presented the high accuracy phase shifts up to 1 GeV are reproduced
very well.  One also finds that the phase shifts for the full model largely
agree to the input phase in the whole energy range considered (this is not the
case for the phase of the form factor, as discussed in the next paragraph) as
well as to the data of Ref.~\cite{hyams,hyams1}~\footnote{Only two of the 4
  solutions presented in that paper are given, for the other two are in strong
  discrepancy with the phase shifts of Ref.~\cite{madrid_new}.}. This happens,
since the resonance couplings to the $\pi\pi$ channel are rather small --- the
resonances show up prominently in the form factor mainly due to the couplings
to the photon. Especially, the present model can not account for the
significant inelasticity visible in the data of Ref.~\cite{hyams1} --- the
according to Ref.~\cite{WOnew} preferred solution $(-+-)$ shows an inelasticity
of the 0.8 already at 1 GeV.
At this point in time it is not possible
to decide whether this failure is an indication of a short coming of the model
used here, or of the data of Refs.~\cite{hyams,hyams1}. What might support the latter
conjecture is that the data on $\eta$ of  Refs.~\cite{hyams,hyams1} are in disagreement
with both the analysis of Ref.~\cite{propo} as well as that of
Ref.~\cite{madrid_new} at around 1 GeV.

In the elastic regime the phase of the form factor has to agree to the phase
of elastic scattering --- a fact known as Watson theorem~\cite{watson}. At higher energies
this connection is lost. In the right panel of  Fig.~\ref{fig:pionFFsmalls}
we show the difference between $\psi$, the phase of the form factor,
and the scattering phase shift $\delta$. Also shown in the panel is the
allowed upper bound of the phase shift difference given in Ref.~\cite{Simon}.
As one can see our amplitudes largely exhaust the range allowed by unitarity.
This reflects again the fact that in the present formalism all resonances
besides the $\rho(770)$ couple to the $\pi\pi$ channel only weakly.

\section{Summary and Outlook}
\label{sec:summary}

In this paper a formalism was presented that allows for a simultaneous
description
of both $\pi\pi$ scattering data as well as form factors
without the need to model the low energy regime: at low energies $\pi\pi$
phases can be used as input directly. At higher energies the formalism maps
smoothly onto the well known $N/D$ method which is similar to the $K$--matrix
approach, however, with improved analytical properties. 
As an example in this paper the formalism was applied to pion pairs in
the $p$-wave. An excellent description is found for the pion vector form
factor in both the neutral channel --- from $e^+e^-$ annihilations, with
$\rho$-$\omega$ and $\rho$-$\phi$ and resonance-$\gamma$ mixing included --- as well as the charged channel --- from
$\tau$ decays. In addition we also found a qualitative agreement with data on
the non 2$\pi$ channels from $e^+e^-$ annihilations --- these data were studied
within a dynamical model here for the first time.

We found, however, that within the given formalism it was not possible to
to describe the behavior of the inelasticity given in  Refs.~\cite{hyams,hyams1}.
 At this point in time we are not able to judge if this deviation
indicates a short coming of the model or points at a problem in the data.
However, the observation that the values of $\eta$ of Refs.~\cite{hyams,hyams1} are
in disagreement with the analyses of Refs.~\cite{madrid_new,propo} at $s\sim
1$ GeV$^2$ might
indicate that there is a problem in the data of  Refs.~\cite{hyams,hyams1}
also at higher energies.

The formalism described here can be applied to all partial waves, especially
also the isoscalar $s$--wave. Here, however, it is less clear what to use for the
elastic phase shift $\tilde \delta$, since the pronounced structure from the $f_0(980)$, which
also couples strongly to $\bar KK$, shows
up already short after the phase reached 90$^0$. We leave this study to a
future work.

\vspace{0.5cm}

\noindent
{\bf Acknowledgment}

\vspace{0.2cm}

\noindent
I thank Maurice~Benayoun, Irinel Caprini, Simon Eidelman, Martin Hoferichter,
Bastian Kubis, Ulf-G.~Mei\ss ner and Juan~M. Nieves for useful and inspiring
discussions and comments to the manuscript and Wolfgang Ochs for useful
remarks about the data of Refs.~\cite{hyams,hyams1}.

\appendix

\section{The two potential formalism}

Let us assume that there is a sensible way to split
the scattering potential into two pieces (as in the
main text the potentials $\tilde V$ and $V_R$, the
vertex function $\Gamma$ as well
as the T-matrices are matrices in channel space, while the
form factor $F$ and the production vertices $M$ are vectors
in channel space)
$$
V=\tilde V+V_R \ .
$$
We will show in this Appendix that the full $T$--matrix can be split
accordingly --- to simplify notations, in the appendix we
do not show the centrifugal barrier factors $\xi$ explicitly. In operator form the Bethe--Salpeter equation
for the $T$ matrix may be written as
$$
T=V+VGT = \tilde V + V_R + (\tilde V+V_R)GT \ .
$$
Here $G_i$ denotes the operator for the integral over
the $n$--particle intermediate state of channel $i$, e.g. for the two--$\pi$
intermediate
state we have
$$
VGV\propto \frac1{i}\int \frac{d^4k}{(2\pi)^4}V(k,..)
\frac1{k^2-m^2+i\epsilon}
\frac1{(k-P)^2-m^2+i\epsilon}V(k,..) \ ,
$$ where $P$ denotes the total 4--momentum of the system, $P^2=s$. Note, not all
arguments of the potential $V$ are shown explicitly.  Introducing $\tilde T$
as the solution of
$$
\tilde T=\tilde V+\tilde V G\tilde T
$$
and the dressed vertex functions 
$$
\Gamma_{\rm out} = 1+\tilde TG \quad \mbox{and} \quad \Gamma_{\rm in}^\dagger = 1+G\tilde T
$$
we get
$$
T=\tilde T+T_R=\tilde T+V_R\Gamma_{\rm in}^\dagger+(V_RG+\tilde VG)T_R \ .
$$
Due to time reversal invariance we have $\Gamma_{\rm out}=\Gamma_{\rm in}^\dagger$.
Since disc($G$)=$2i\sigma$ and disc$(\tilde T)=2i\sigma \tilde T^*\tilde T$ one has
$$
\mbox{disc}(\Gamma_{\rm out}(s))=\mbox{disc}(\tilde T)G+\tilde
T^*\mbox{disc}(G)=\sigma \tilde T(s)^*\Gamma_{\rm out}(s) \ ,
$$
thus $\Gamma_{\rm out}$ holds the  unitarity relation for a form factor
in a channel where the interactions are given by $\tilde T$.

We may therefore define
$$
T_R=\Gamma_{\rm out}t_R \Gamma_{\rm in}^\dagger
$$
and derive with $\Sigma=G\Gamma_{\rm out}$ and $\tilde VG\Gamma_{\rm out}=\Gamma_{\rm out}-1$
$$
t_R=V_R+V_R\Sigma t_R \quad \longrightarrow \quad t_R = \left[1-V_R\Sigma \right]^{-1} V_R \ .
$$
From the definition above the discontinuity of the self--energy $\Sigma$ is found
to be
$$
\mbox{disc}(\Sigma)=\mbox{disc}(G)\Gamma_{\rm out}+G^*\mbox{disc}(\Gamma_{\rm out})
=2i\underbrace{\left(1+G^*\tilde T^*\right)}_{\Gamma_{\rm
    out}^*}\sigma\Gamma_{\rm out} \ ,
$$
which was used to derive Eq.~(\ref{Ldef}).


\end{document}